\documentclass[aps,twocolumn,prl]{revtex4}
\usepackage{bm}
\usepackage{graphicx}
\usepackage{amsbsy}
\usepackage{amsmath}
\usepackage{amsfonts}

\newcommand{\ket}[1]{|{#1}\rangle}
\newcommand{\nn}{\nonumber}
\newcommand{\dg}{^\dagger}

\newcommand{\per}{\text{per}}

\begin{document}

\title{Producing a pure single-photon state from pure superposition states}

\author{Dominic W.\ Berry} 
\affiliation{Department of Physics, The University of Queensland, Brisbane,
Queensland 4072, Australia}

\begin{abstract}
Two pure states in superpositions of zero and one photons may be processed,
via beam splitters and photodetection, to yield a pure single-photon state.
\end{abstract}

\date{\today}

\maketitle

One of the requirements of linear optical quantum computation \cite{KLM}
is the availability of single-photon sources. As available single-photon
sources do not produce pure single-photon states, it would be desirable to
postprocess imperfect single-photon sources to produce a state closer to a
pure single-photon state. This approach was examined in
Refs.\ \cite{berry1,berry2,Myers}. In Ref.\ \cite{berry2}, it was shown that
if pure superposition states were available, then it would be possible to
produce a pure single-photon output state from three of these superposition
states. Here I present an improved scheme that requires only two
superposition states, and has a higher probability for success.

\begin{figure}
\centerline{\includegraphics[width=2cm]{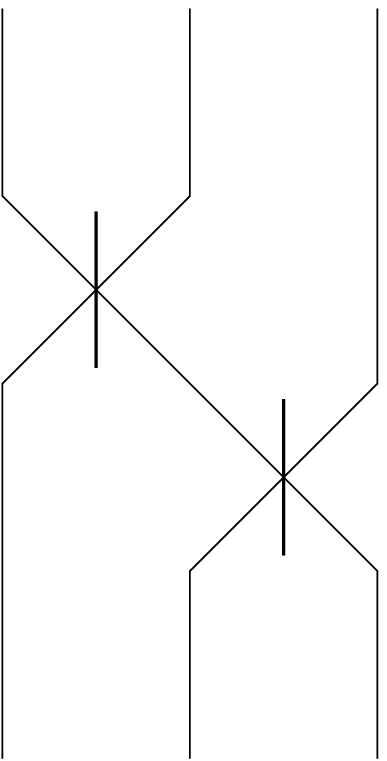}}
\begin{picture}(0,0)
\put(-33,2){$|0\rangle$}
\put(-6,2){$|\psi_{p_1} \rangle$}
\put(22,2){$|\psi_{p_2} \rangle$}
\put(-35,126){$|\psi \rangle_{\text{out}}$}
\put(-2.5,126){1}
\put(25.5,126){0}
\put(-20,58){${\bm \Lambda'}$}
\put(10,66){${\bm \Lambda}$}
\end{picture}
\caption{The interferometer for obtaining a pure single-photon state from two
pure superposition states.}
\end{figure}

Consider two input modes that are in the states
\begin{equation}
\ket{\psi_{p_1}} = \alpha_1\ket{0}+\beta_1\ket{1}, \quad
\ket{\psi_{p_2}} = \alpha_2\ket{0}+\beta_2\ket{1},
\end{equation}
where $p_k=|\beta_k|^2$ is the probability for a single photon. For
additional generality, the input states can be non-identical.
The initial state may be written as
\begin{equation}
\ket{\psi}_{\text{in}}^{(2)} = \left[\alpha_1\alpha_2+\alpha_2\beta_1 a_1\dg +
\alpha_1\beta_2 a_2\dg + \beta_1\beta_2 a_1\dg a_2\dg \right] \ket{00} .
\end{equation}
Applying the beam splitter transformation gives
\begin{align}
\ket{\psi}_{\text{trans}}^{(2)} & = \left\{\alpha_1\alpha_2+\alpha_2\beta_1
(\Lambda_{11}a_1\dg+\Lambda_{21}a_2\dg)  \right. \nn \\ & \quad \left. +
\alpha_1\beta_2(\Lambda_{12}a_1\dg+\Lambda_{22}a_2\dg) + \beta_1\beta_2
\left[\Lambda_{11}\Lambda_{12}(a_1\dg)^2 \right.\right.
\nn \\ & \quad \left.\left. + \Lambda_{21}\Lambda_{22}
(a_2\dg)^2 + (\per\bm{\Lambda}) a_1\dg a_2\dg \right]\right\} \ket{00} .
\end{align}
Conditioning on detection of zero photons in mode 2 gives the output state
\begin{align}
\ket{\psi}_{\text{out}}^{(2)} &\propto \alpha_1\alpha_2\ket{0} +(\alpha_2\beta_1
\Lambda_{11} + \alpha_1\beta_2\Lambda_{12})\ket{1} \nn \\
 & \quad + \sqrt{2}\beta_1\beta_2\Lambda_{11}\Lambda_{12}\ket{2}.
\end{align}

By taking the appropriate beam splitter parameters we can remove the single
photon contribution entirely, giving
\begin{equation}
\ket{\psi}_{\text{out}}^{(2)} \propto \alpha_1\alpha_2\ket{0}
+ \sqrt{2}\beta_1\beta_2\Lambda_{11}\Lambda_{12}\ket{2}.
\end{equation}
It is clear that, if we combine this state with the vacuum, and condition
upon detection of a single photon, the remaining mode must have a single
photon. In the case that the inputs are identical, a 50/50 beam splitter is
required, and the probability for success is $|\beta|^4/4$. This probability
is a significant improvement over that for the scheme given in
Ref.\ \cite{berry2}, which was $16|\beta|^6/81$.

This scheme is more promising for experiments, as it only requires two
superposition states, and the probability of success is higher. Although
single-photon sources do not produce pure superposition states, these states
can be produced by quantum scissors \cite{Pegg,Babichev}. The quantum scissors
scheme requires a single-photon input, so this method is not practical as a
means of producing single photons. However, this is a possible method of
performing proof of principle experiments.

I would like to thank B.\ C.\ Sanders and C.\ R.\ Myers for valuable discussions.

\end{document}